\documentclass[prc,aps,floatfix,nofootinbib,preprint]{revtex4}
\usepackage{graphicx}   
\usepackage{amssymb}
\usepackage{amsmath}
\usepackage{amsfonts}

\def\be{\begin{equation}}
\def\ee{\end{equation}}

\def\ca{$^{40}$Ca}
\def\zr{$^{80}$Zr}
\def\u{$^{236}$U}

\begin{document}

\title{
The shapes of nuclei}
 
\author{G.F.~Bertsch\footnote{bertsch@uw.edu}}
 
\affiliation{
Department of Physics and Institute for Nuclear Theory, 
\\ University of Washington, Seattle, Washington 98915, USA}
 
\begin{abstract} 
Gerry Brown initiated some early studies on the coexistence of different
nuclear shapes.  The subject has continued to be of interest and is
crucial for understanding nuclear fission. We now have a very good
picture of the potential energy surface with respect to shape degrees
of freedom in heavy nuclei, but the dynamics remain
problematic.  In contrast, the early studies on light
nuclei were quite successful in describing
the mixing between shapes.  Perhaps a new approach in the
spirit of the old calculations could better elucidate the character
of the fission dynamics and explain phenomena that
current theory does not model well.
\end{abstract} 

\maketitle
 
\section{Introduction}

In the early 1960's, Bohr and Mottelson pointed out some puzzling
experimental data:  some light nuclei thought to be
spherical in shape had excited energy levels exhibiting characteristics
of deformed nuclei.  This was taken up first by Engeland \cite{en65} and
then
Gerry Brown, who saw an opportunity to test the realistic nuclear
interactions
that were being developed at the time. The nuclei $^{16}$O and $^{18}$O were
the first
subjects of study \cite{br66}.  As his graduate student, I worked on
a parallel study of Ca isotopes as part of my thesis project.  Later,
a more definitive study of the Ca nuclei was carried out by
Gerace and Green \cite{ge67}.  As a general conclusion, one saw that
the mixing between shapes could be understood with
the realistic interactions derived from nucleon-nucleon
scattering data.

Since those early days of nuclear structure physics, the subject
of nuclear deformation has matured.
First of all, we now know that the shape coexistence is ubiquitous
in the low-energy spectra of nuclei across the periodic table, affecting
even the fission properties of the heaviest nuclei.  Also, we now
have computational tools to describe and predict the static features of
the landscape of nuclear shapes.  However, the dynamics of shape change, ie.
how different shapes mix together,  has been a
challenging problem in the theory of heavy nuclei and is still not well
understood.  I shall describe some
work I have been engaged in recently, to develop a new approach
to fission dynamics in the spirit of the old studies on
light nuclei.

\section{Theory of static deformations}
Nuclei are highly deformable, and the first task is to construct reliable
models of the nuclear potential energy surface, ie. the energy of 
configurations as a function of deformation coordinates such as the
expectation values of quadrupole and higher moments.  An immediate
question is how to define both energy and shape of a configuration:
the operators for these two quantities do not commute.  The resolution
of this conundrum is that we are only dealing with approximate wave
functions, not the true eigenstates of the Hamiltonian.
In practice, theory relies on the mean-field representation of 
wave functions as products of single-particle orbitals.  
The present-day calculational framework is very similar to
the Density Functional Theory (DFT) of condensed matter physics.
One defines an effective interaction,  which may depend on the local
density.  The energies and shapes are determined by minimizing the
energy expectation value just as in Hartree-Fock (HF) theory.

\begin{figure}[tb] 
\begin{center} 
\includegraphics[width=10 cm]{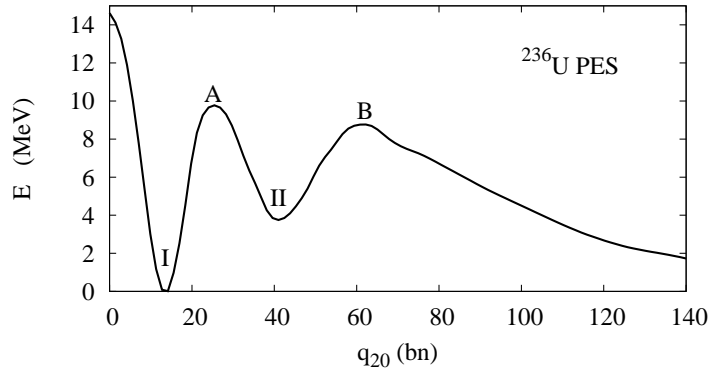} 
\caption{Potential energy surface for $^{236}$U covering the range
of deformations up to the scission point.  The curve was calculated
by minimizing the D1S density functional with the mass quadrupole
moment constrained to the value on the abscissa. The point marked ``I" is
the global minimum.  The points marked ``A" and ``B" are saddle points
on the fission path, with a second minimum ``II" in between.
}
\label{PES} 
\end{center} 
\end{figure}

Several families of DFT in use today, and all have considerable
predictive power on static properties of nuclei.  In the examples discussed
below, I will show results obtained with the Gogny D1S functional.
\cite{de10}.  It has 14 parameters, 3 of them fixed and 11 adjusted
many years ago \cite{D1S} to
reproduce  general nuclear properties.  An example showing its
considerable predictive power is 
recent systematic study
of the low-energy spectroscopy of even-even nuclei \cite{de10}.  I will
come back to findings from this study later.  Another success of the
DFT approach is verification of additional shape minima at very high
deformation in heavy nuclei.  The potential energy surface that
a fissioning nucleus traverses has at least two minima and perhaps
more.  Fig. \ref{PES} shows the energy versus deformation in a
typical fissile nucleus.
\section{How nuclei change shape}
Up to now, the only practical approach for treating shape dynamics
in heavy nuclei is the Generator Coordinate Method (GCM) proposed by 
Hill and Wheeler \cite{hi53}.  
Formally, one generates
a continuum of configurations by minimizing the Hamiltonian in the 
presence of an external field.  
Thus, the configuration $|\Psi_\lambda\rangle$ is defined by
minimizing
\be
\langle \Psi_\lambda | \hat H - \lambda \hat Q | \Psi_\lambda\rangle
\ee
where $\hat H$ is the Hamiltonian or energy-density functional and 
$\hat Q$ is a one-body operator such as the axial quadrupole field
\be
\hat Q_{20} = \hat z^2 -(\hat x^2+\hat y^2)/2.
\ee
Then one constructs an effective Hamiltonian from information about
the diagonal and off-diagonal matrix elements  
$
\langle \Psi_\lambda |   \Psi_\lambda'\rangle
$ and 
$
\langle \Psi_\lambda | \hat H | \Psi_\lambda'\rangle
$.

Computationally, the GCM can be implemented
in two ways.  In the first way, the moment $q = 
\langle \Psi_\lambda | \hat Q | \Psi_\lambda\rangle$
is treated as the position variable 
in a one-dimensional Schr\"odinger equation.  It is straightforward to
derive formulas for potential energy function $V(q)$ and a kinetic
energy operator 
\be
-{1\over 2}{\partial \over\partial q} { 1\over {\cal I}(q)} {\partial\over \partial q}
\ee
in approximations such as the Gaussian
Overlap Approximation \cite{gi83}.  This one-dimensional collective
approach works well
for treating the ground state tunneling under the barriers, but generalizing
it to include excited states is cumbersome \cite{be11}.
A more technical 
problem is necessity to define at least four or five deformation operators
to completely specify the shape of the fissioning nucleus.  It is 
possible, but not easy, to set up and solve the corresponding
higher-dimension Schr\"odinger equation; a two-dimensional approximation
was carried out in Ref. \cite{go05}.

The second computational method
is to discretize the GCM configurations with a mesh of $\lambda$ values
and minimize the Hamiltonian in that finite basis.  There are obstacles to this approach as well.
On a purely technical level, the 
fact that the configurations are not orthogonal leads to numerical
instabilities.  More serious on a fundamental level, it is not
clear how to calculate interaction matrix elements between configurations
when they are constructed via an energy density functional.  Several
plausible
prescriptions are possible, but the results can be unphysical if the 
wrong prescription used \cite{ro10}.  Such problems never occurred
in the early work on light nuclei---we used orthogonal bases and
we had a pretty good idea of the interactions.

To build an alternative to the GCM approach, we should start 
by 
constructing an orthonormal basis within the mean-field framework.
The dynamics can be developed later from the off-diagonal interactions
in this basis.  In the old work, the basis was constructed from the
harmonic oscillator Hamiltonian by using the associated SU(3) group structure
to organize the many-body configurations.  This is obviously too crude
for treating heavy nuclei, and it is better to use the DFT to build the
orbitals.  
In the nuclear
context, one could use DFT orbitals and still preserve orthogonality
by to constraining them by
their quantum numbers rather than by their deformations.  
For most nuclei, the mean-field potential is  axially symmetric and 
has good parity at the DFT minima.  This allows one to assign 
orbital quantum numbers $K^\pi$,
where $K$ is the angular momentum about the symmetry axis and $\pi$ is
the parity of the orbital.  The DFT minimization can be carried out
taking as a constraint the number of nucleons having given values of $K$, $\pi$,
and isospin projection $\tau_z$.  Wave functions that differ in filling
numbers are automatically orthogonal, so this should be very helpful
for constructing an orthogonal basis.  In the remaining sections of this article,
I will go through some examples illustrating the use of a $K^\pi$
partitioning to define the configurations.

\begin{figure}[tb] 
\begin{center} 
\includegraphics[width=10 cm]{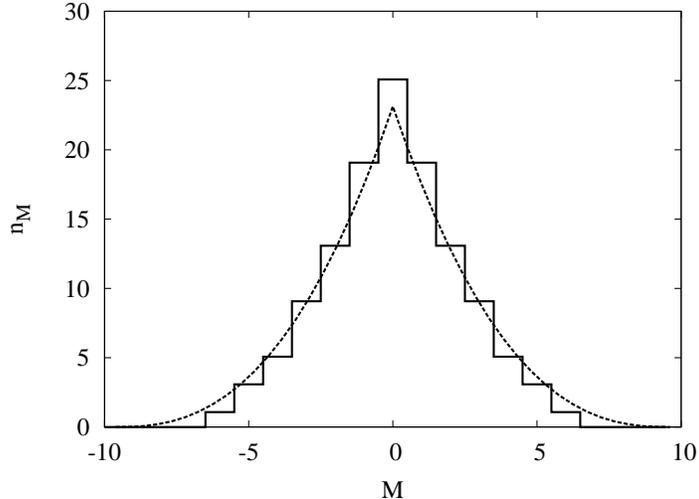} 
\caption{$M$ distribution of occupied neutron orbitals in the
nucleus $^{208}$Pb.  Histogram:  spherical shell model. 
Dashed line  Eq. (\ref{eq-semiclassical} with
$R=7.1$ fm, $k_F= 1.35$ fm$^{-1}$.  
}
\label{semiclassical} 
\end{center} 
\end{figure}

Before going on to the examples, it is instructive to see how the
partitioning works in a semiclassical limit.  Assume that the
nucleus is spherical and the phase space density is uniform in 
spheres of radius $R$ in coordinate space and 
$k_F$ in momentum space.  Then the number of nucleons having
orbital quantum number $M$ is given by
\be
n_M = g\int^R d^3 r \int^{k_F} { d^3 k \over (2 \pi)^3 }\delta(
M - (\vec r \times \vec k)_z)
\ee
where $g=2$ is the spin degeneracy of the nucleon and $M$ is
the orbital angular momentum about the $z$-axis.  The integrals 
can be carried out analytically; the result is
\be
n_M = {9 N \over 4k_F R} \left( (1/2 + x^2) {\rm arccos}\, x - 3 x
(1-x^2)^{1/2}\right)
\label{eq-semiclassical}
\ee
where $x = M/(k_F R)$.  The same formula applies to ellipsoidally
deformed nuclei with the radius  $R$ replace by the transverse 
radius of the ellipsoid.
Fig. \ref{semiclassical} shows the distribution for the
neutrons in the spherical nucleus $^{208}$Pb, comparing the mean-field 
fillings with the formula (\ref{eq-semiclassical}).

\section{Examples} 
The shells of the three-dimensional harmonic oscillator model are
completely 
filled or empty at nucleon numbers 8, 20, 40,...  The first two examples 
here are nuclei at those possibly magic numbers:  \ca~ and \zr.  The third
example deals with the early shape changes in a fissioning \u~nucleus. 
\subsection{\ca}
\begin{figure}[tb] 
\begin{center} 
\includegraphics[width=8cm]{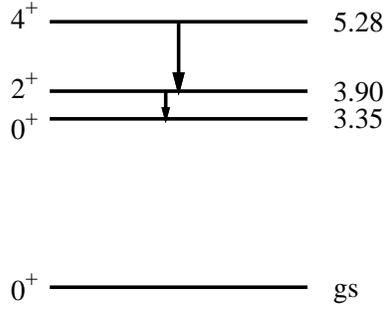} 
\caption{The spectrum of \ca, showing the first three levels of the
deformed band and the gamma transitions that establish the deformation
of the band \cite{ma71}}
\label{ca40-levels} 
\end{center} 
\end{figure}

The spectroscopy of \ca~has been known since the early 1960's.  
Fig. \ref{ca40-levels} shows the states in the spectrum that are 
relevant to the discussion.
The first excited state in the spectrum has angular moment zero, 
which is rare among the 600 or so known even-even nuclei.  
The excited $0^+$ together with the  
$2^+$ and $4^+$ above it form the lowest members of a rotational band.
The evidence that these states are part of a band
comes from the energy spacing and from the electromagnetic transition rates
between states, indicated by arrows in the Figure.  The measured quadrupole 
transition strength between two lowest states in the band is 
$B(E2, 2^+_1\!\rightarrow\!{0^+_2}) = 250 \pm 35 $ e$^2$ fm$^4$ \cite{ma71}.
On the scale of a
single-particle quadrupole moment this is very large and  would require an axis ratio
of $a_z/a_\perp \approx 1.55  $ to model the band 
as an ellipsoidal rotor.  The corresponding intrinsic mass quadrupole
moment in the band is 
\be
\langle\, \hat Q_{20} \,\rangle = 117~~{\rm  fm}^2.
\label{q-ca}
\ee

\begin{figure}[tb] 
\begin{center} 
\includegraphics[width=10cm]{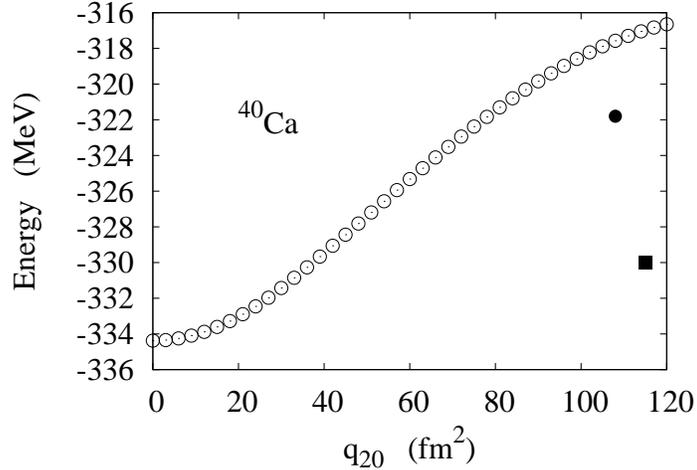}
\caption{Energies of \ca~configurations calculated by the GCM (open circles)
and by  $K^\pi$-constrained minimization (black circle).  
}
\label{E-vs-q} 
\end{center} 
\end{figure}

\begin{figure}[tb] 
\begin{center} 
\includegraphics[width=4cm]{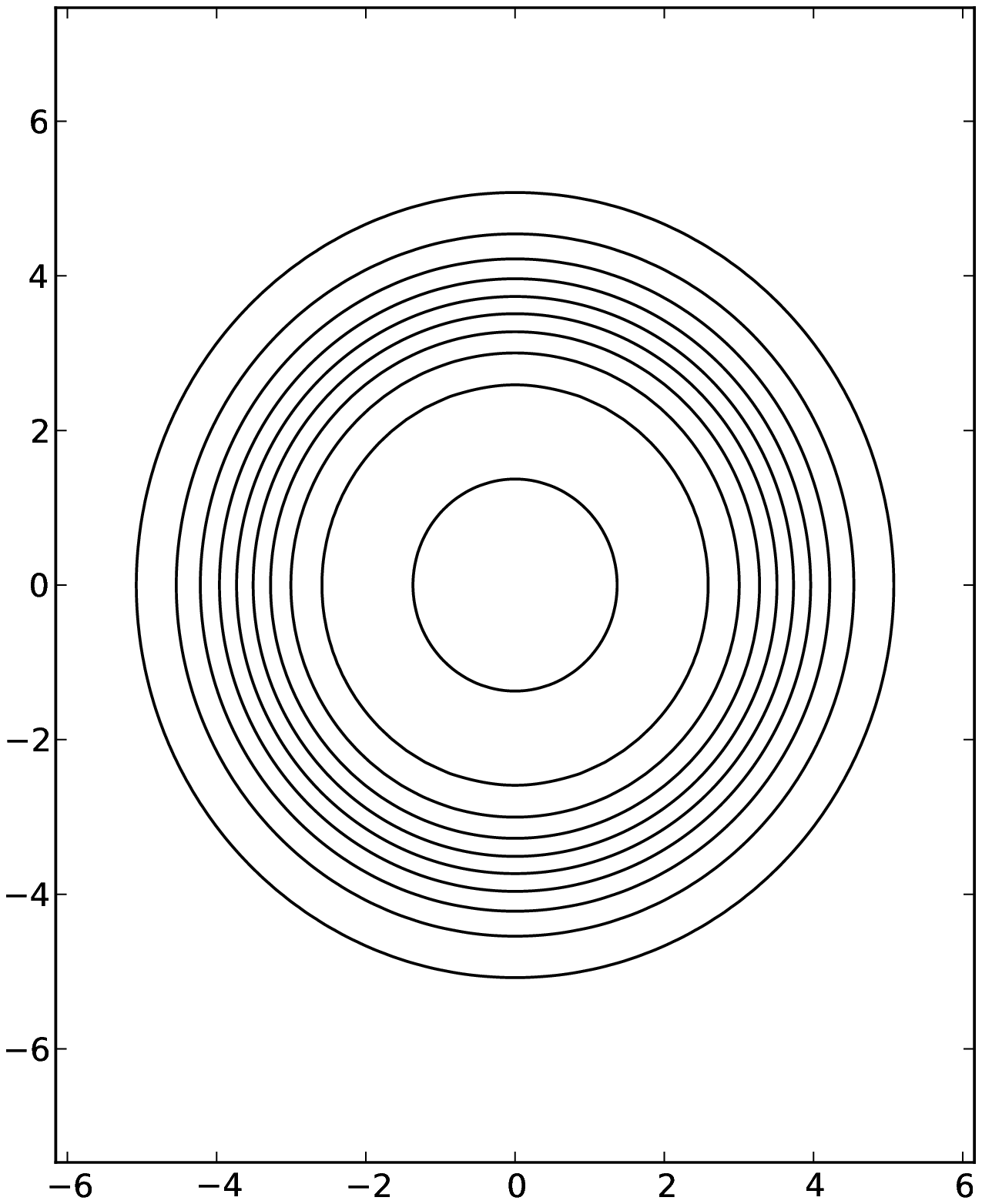}%
\includegraphics[width=4cm]{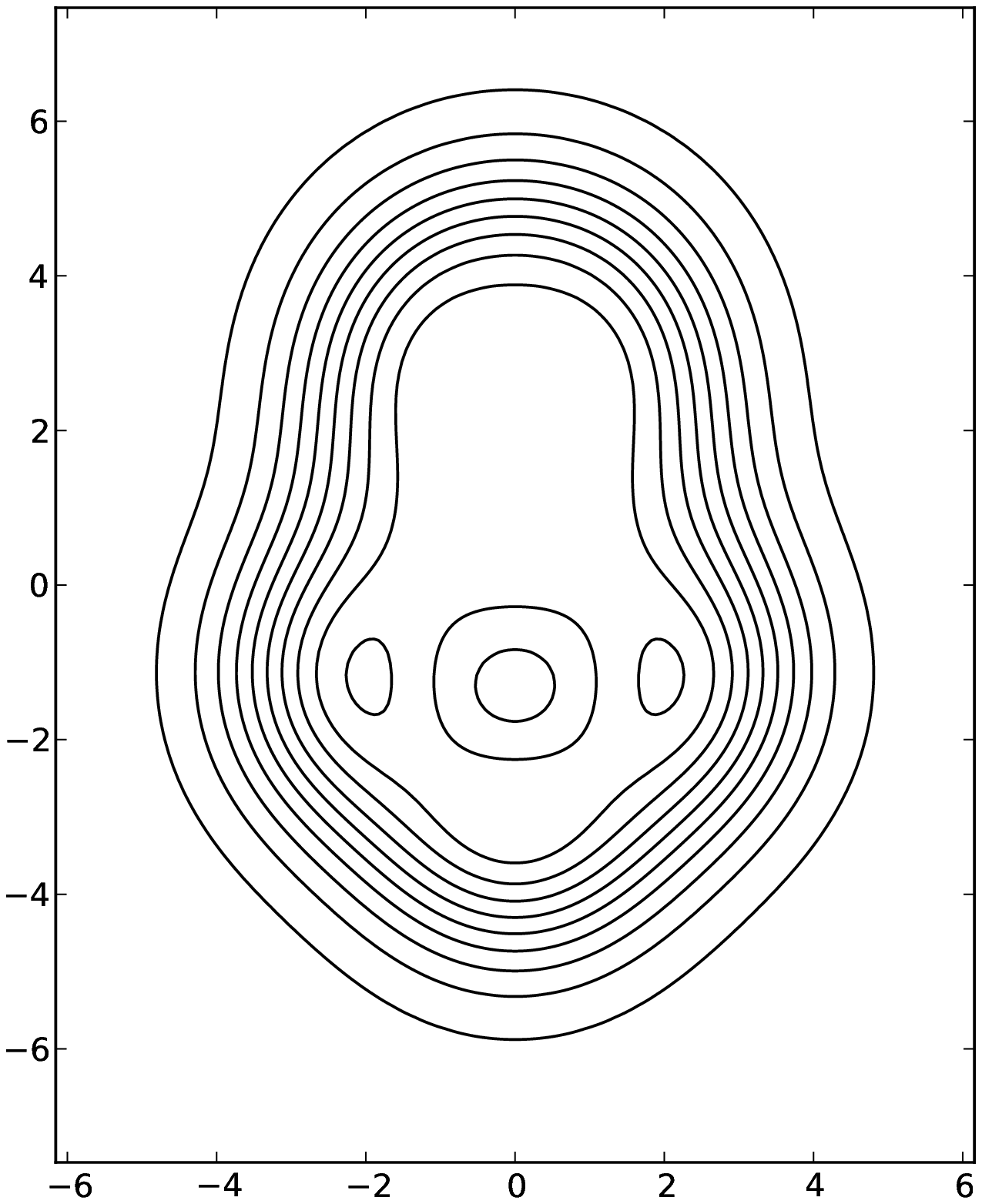}%
\includegraphics[width=4cm]{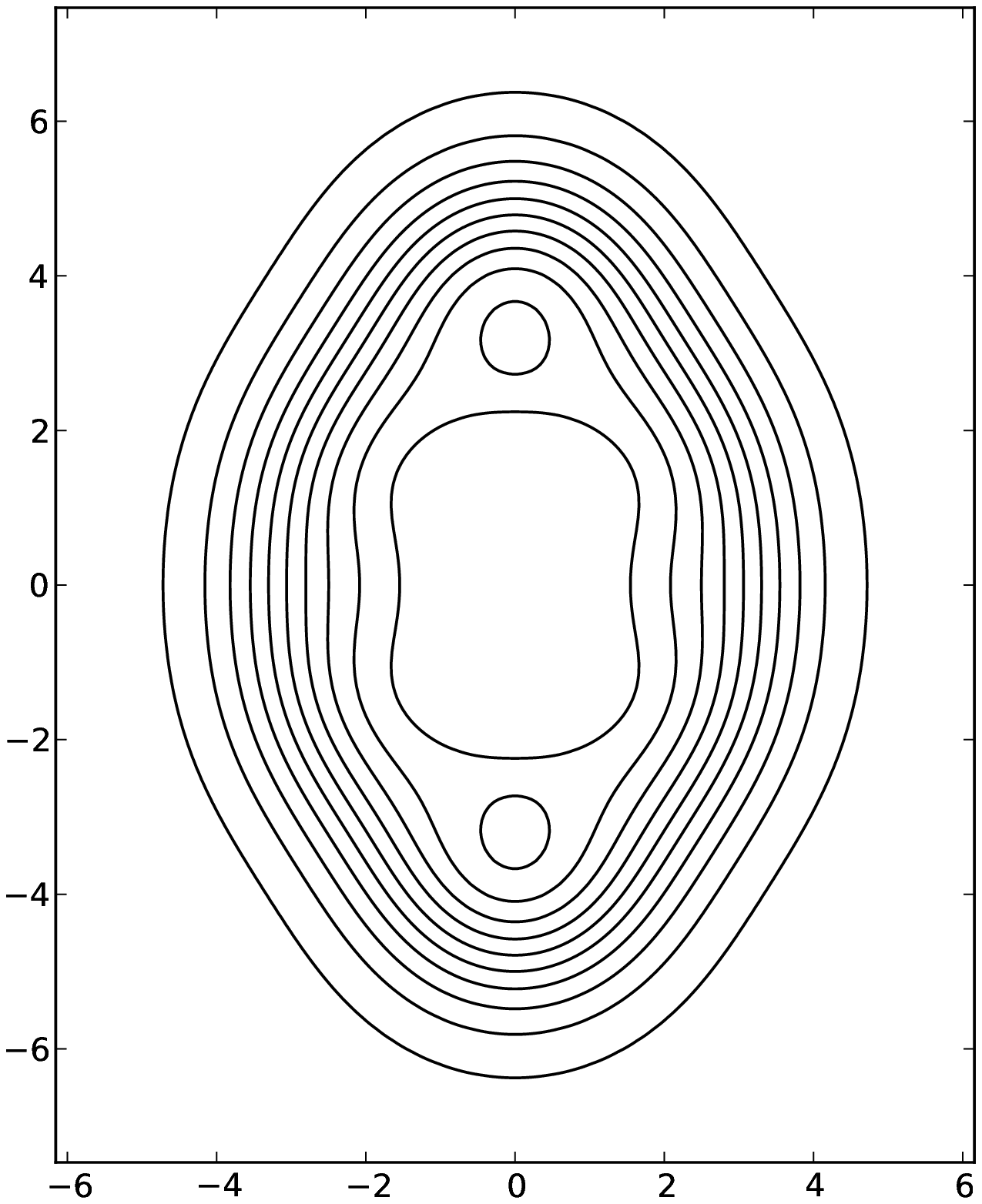}%
\caption{Density distribution of the \ca~configurations.  Left:
ground state; center: GCM configuration at $q = 108$ fm$^2$; 
right:  4p-4h $K^\pi$-constrained configuration.
}
\label{ca-shapes} 
\end{center} 
\end{figure}

Now let us compare the GCM and the $K^\pi$-partition approaches to 
determine the structure of the band.
We start with the GCM, applying a
field $\lambda \hat Q_{20}$ to the spherical ground state configuration.  
The minimization is performed
small increments of $\lambda$ up to the point where the configuration has
the deformation (\ref{q-ca}) extracted from experiment. The results are shown
in Fig. \ref{E-vs-q} as the open circles.
The energy increases monotonically as the
deformation increases.  At the deformation corresponding to the
one extracted from experiment, the excitation energy
is 13 MeV.  This is way off from the observed 
band head at 3.35 MeV. There will be a
small gain of energy when one takes into account that the calculated
energy is that of the band as a whole, but the energy gain is
probably too small to approach the experimental value. 
Anyway, a big surprise comes when we look at the shape of the
nucleus, shown as in the middle panel of Fig. \ref{ca-shapes}.  
One sees that
the GCM-generated configuration no longer has good parity;
there is a strong octupole component as well as the quadrupole
deformation.  This would imply that the band should have odd-parity members
interleaved between the even-parity members.  Since
that is not the case, so we can concluded that GCM carried out this way
has failed.

\begin{figure}[tb] 
\begin{center} 
\includegraphics[width=10cm]{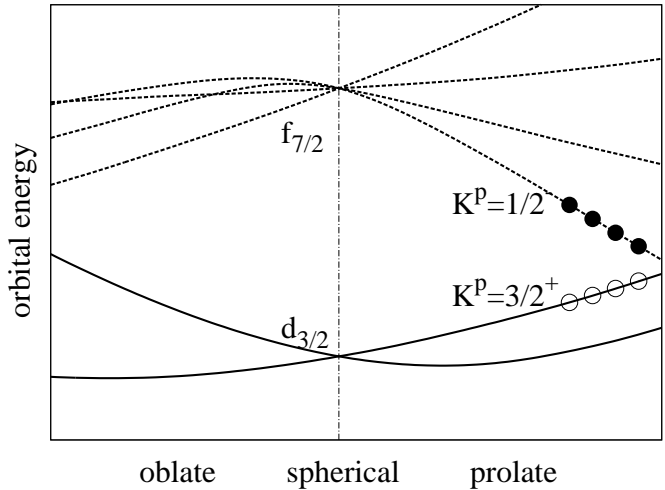}
\caption{Single-particle energies of orbitals around the Fermi level
of \ca~as a function of deformation.  The deformed band is
attributed to 4-particle 4-hole excitation shown with solid and
open circles (after Fig. 1 of Ref. \cite{ge67}.}
\label{4p4h} 
\end{center} 
\end{figure}

Now let's try the $K^\pi$-constrained approach using the $K^\pi$ partitions from
Ref. \cite{ge67}.  The ground state configuration has the fillings of the
spherical shell model.  The deformed configuration was built
by taking the lowest 4-particle 4-hole
excited state in a deformed harmonic oscillator potential.  This
is the Nilsson model; its diagram of orbital energies 
is shown in Fig. \ref{4p4h}.   We carry out the
DFT minimization again, but now use the $K^\pi$ quantum numbers of
the occupied orbitals to constrain the minimization. The results
for the spherical and deformed configurations are shown in 
Fig. \ref{E-vs-q} as the
black square and black circle, respectively.  The predicted 
quadrupole moment of the deformed configuration, 105 fm$^2$, agrees 
well with (\ref{q-ca}).  The energy is still too high, but it is lower
than the GCM energy and so is a better candidate for understanding
the band structure.
Of course, one can obtain the configuration by GCM minimization, but
doing this would require a different starting point or 
additional shape-dependent constraints.  
One final point: as may be seen in the right-hand panel of
Fig.~\ref{ca-shapes},
the configuration found by starting from the 4p-4h hole configuration
preserves the even parity of the band.

Ignoring the energy problem, one can try to calculate the mixing
of deformed and spherical states, as was done by the early researchers.
Unfortunately, as I mentioned earlier,
there is no consistent way to extract configuration-interaction elements 
from a DFT.  But DFT should still be 
useful to construct the configurations.  The Hamiltonian matrix elements
might be evaluated using these configurations as in the condensed-matter
hybrid procedure.

\subsection{\zr~region}
\begin{figure}[tb] 
\begin{center} 
\includegraphics[width=10 cm]{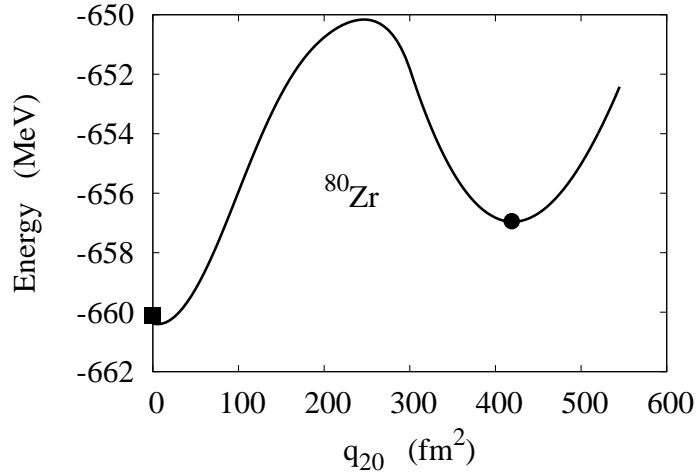} 
\caption{The \zr~potential energy surface. The solid line is generated
by the GCM procedure, starting from a octupole-deformed ground state
minimum.  The $K^\pi$-constrained minimization starting from the partition
of the spherical shell model configuration is shown at the black square.
Starting with the 12p-12h partitions gives the filled circle, coinciding
with the second minimum of the GCM potential energy curve.
}
\label{zr80} 
\end{center} 
\end{figure}

At the next harmonic oscillator shell closure ($N = 20$ at \zr), the
competition between spherical and deformed configurations plays out 
differently.  
The coexistence question here was addressed in the DFT 
by Zheng and
Zamick  \cite{zh91}.  They tried several energy functionals in the
Skyrme family and found that a deformed configuration came out lower than the
spherical.  While in \ca~there were 4 particle jumps needed to connect
the configurations, 12 jumps were needed in \zr.
The Gogny D1S functional yields rather similar results. The corresponding
potential energy surface 
is shown in Fig. \ref{zr80}.  The filled square and circle show the minima for
the spherical and deformed configurations, obtained by constraining
the $K^\pi$ partitions according to the spherical shell model and the
12p-12h Nilsson configuration.  
The latter has a very large
deformation, $Q_{20}\approx 400$
fm$^2$, at an energy $~3.5 $ MeV above the global minimum.  
The deformation is very robust with respect to 
the choice of energy functional.  In terms of the dimensionless
deformation parameter $\beta$, the Skyrme functionals 
and the Gogny D1S all give $\beta = 0.44\pm 0.01$.  The authors
of Ref. \cite{zh91} also note that a simple formula derived earlier by 
myself is quite accurate.  

The energetics are more complicated.
First of all, the shell model configuration is not the lowest energy
minimum at $q= 0$.
The actual minimum has broken parity with a large octupole moment and a
very small quadrupole moment.  The entire potential energy surface
up to high deformation can be generated by 
GCM procedure, provided one starts with a configuration that 
already has an octupole deformation.  It is shown as the solid line
in the Figure. 

\begin{figure}[tb] 
\begin{center} 
\includegraphics[width=11 cm]{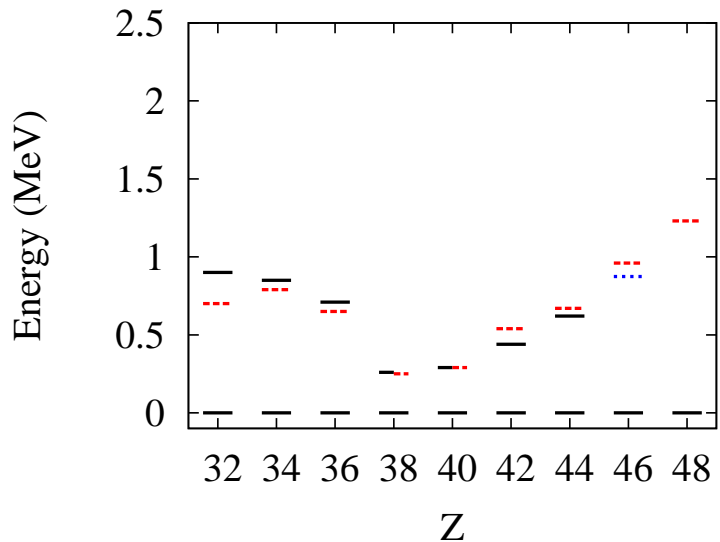} 
\caption{Systematics of $2^+$ excitations in the heaviest known $N=Z$
nuclei.  Solid black lines: older experiments; red dashed line: Gogny/GCM theory
\cite{de10}.  Theory and experiment are visually indistinguishable for the
nuclei $^{76}$Sr and \zr.  The blue dotted line shows the new measurement
at $^{96}$Pd \cite{ce11}.
}
\label{2plus-systematics} 
\end{center} 
\end{figure}

While the 12p-12h is not the lowest configuration in the Gogny DFT, correlation
effects can change the ordering.  Recently \cite{de10} a systematic study of
low-energy spectroscopy was carried out including all five 
quadrupole shape degrees of freedom in
the GCM framework. In that way, rotational energies and some shape 
mixing effects are taken into account. As an example of the predictive power
of the Gogny functional and the method, it was found that calculated 
quadrupole transition moments of deformed nuclei agree with experiment 
to $~10\%$ accuracy. 
For the \zr~nucleus, the authors  found that 
the rotational energy was enough to bring
the band head of the highly deformed configuration down to the
ground state.    Fig. \ref{2plus-systematics}
compares their calculated $2^+$ excitation energies across the chain
of the heavy $N=Z$ even-even nuclei. One sees that $2^+$ excitation
energy in \zr~agrees very well with experiment.  It is also the most
highly deformed in the chain, judging by the 
excitation energy of the lowest $2^+$ state.  
The heaviest measured nucleus in the Figure, $^{96}$Pd, is
a real prediction, as the experimental measurement \cite{ce11} was reported 
the following year.   So, despite the mixed performance of
DFT in \ca, we find that the energy functionals 
becomes quite successful in a heavier region of nuclei.
\section{Fission dynamics}
My ultimate goal  is to gain a better understanding of
the dynamics of fissioning nuclei.  As mentioned at the beginning,
the GCM approach works well for describing  spontaneous fission as
tunneling through a barrier in the potential energy surface.  But
when the excitation energy is above the 
barrier, it is far from clear what theoretical approach is 
justified.  In any case, the spontaneous fission show that 
the part of the interaction responsible for pairing is very important \cite{ro14}.  
This
suggests that the time-dependent Hartree-Fock-Bogoliubov approximation
(TD-HFB) might be appropriate for above-barrier dynamics.  Recently,
computer power has become available to test the TD-HFB for 
fission using current nuclear energy functionals.  Such a
study was carried out by Bulgac and collaborators \cite{bu16}.
Their starting point was a very deformed configuration 
past the second barrier
and at a small excitation energy over the potential energy surface.
They found that fission would take place, but the duration before the
fragment formation could be very long.

A weak point of the TD-HFB approximation is that it takes into
account only a very restricted set of interaction matrix elements
to propagate the system
from one configuration to another.  Namely, configurations that
can be generated as two-quasiparticle excitations of the starting
configuration are included in some way, but all four-quasiparticle
transitions are neglected.  In the example of configuration mixing
in $^{40}$Ca, the HFB approximation would be useless.  There is
no pairing condensate, and even if there were one, the intermediate
configurations are only partially represented by the allowed pair
jumps.  

\begin{figure}[tb] 
\begin{center} 
\includegraphics[width=10cm]{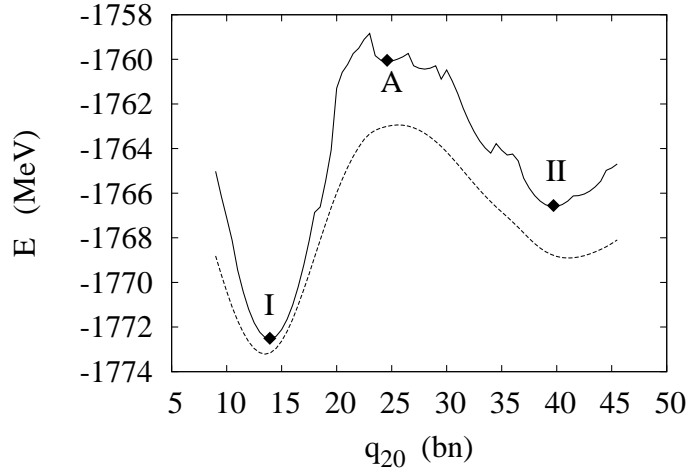} 
\caption{\u~HFB potential energy surface between the first and second
minima and the energies of the partition-constrained HF wave
functions.  Solid and dashed lines are the HF and HFB potential energy 
surfaces respective.  Diamonds show the energy and quadrupole moment of
the partition-constrained minima.  
}
\label{u236-pes} 
\end{center} 
\end{figure}

While it would still be an enormous challenge to make a realistic
treatment of fission in the discrete-configuration approach, we
can at least see how the landscape could be traversed.  We concentrate
on the two minima and the barrier between them.
Fig. \ref{u236-pes} shows
the potential energy surface of \u~again, expanding the 
horizontal scale in the
region of the first barrier.  
%
%
\begin{figure}
\centerline{
\includegraphics[width=8cm]{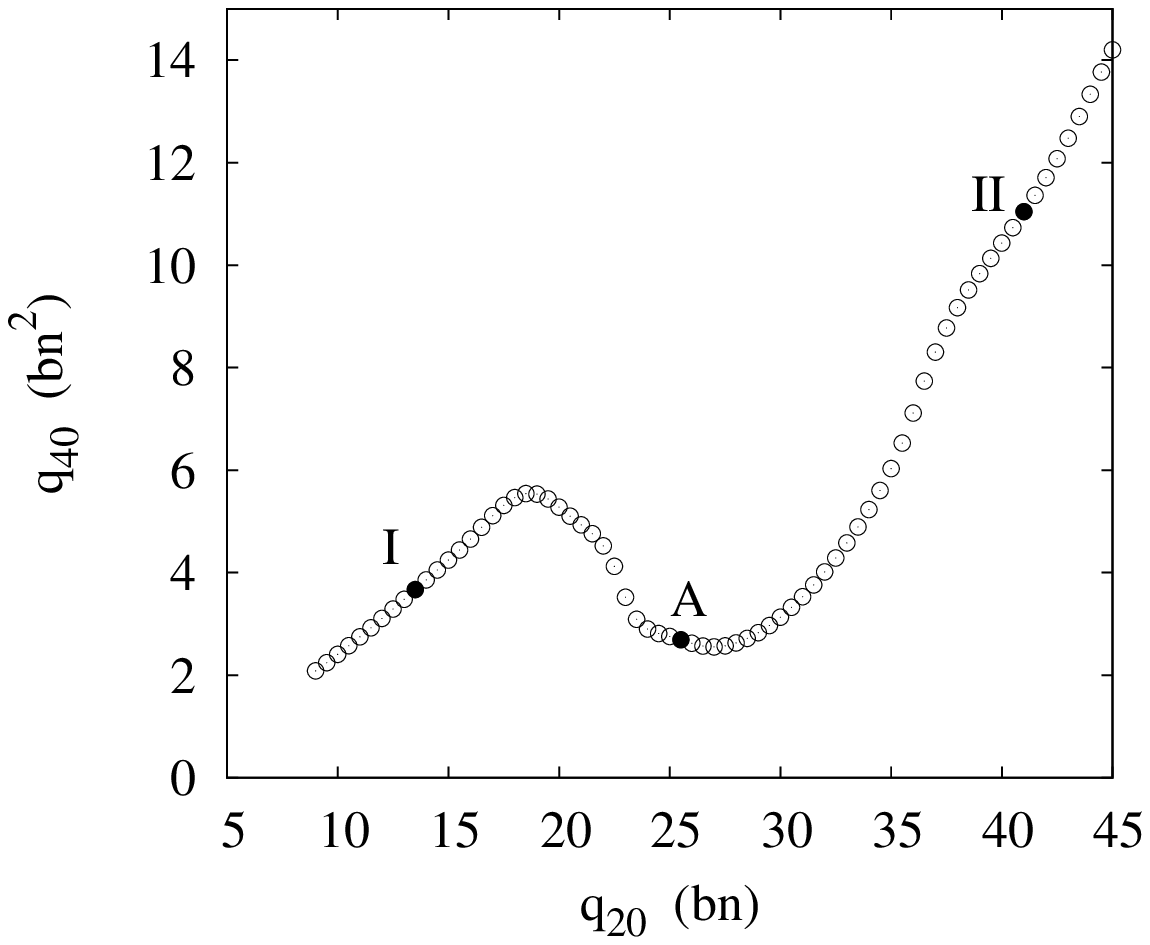}
\includegraphics[width=8cm]{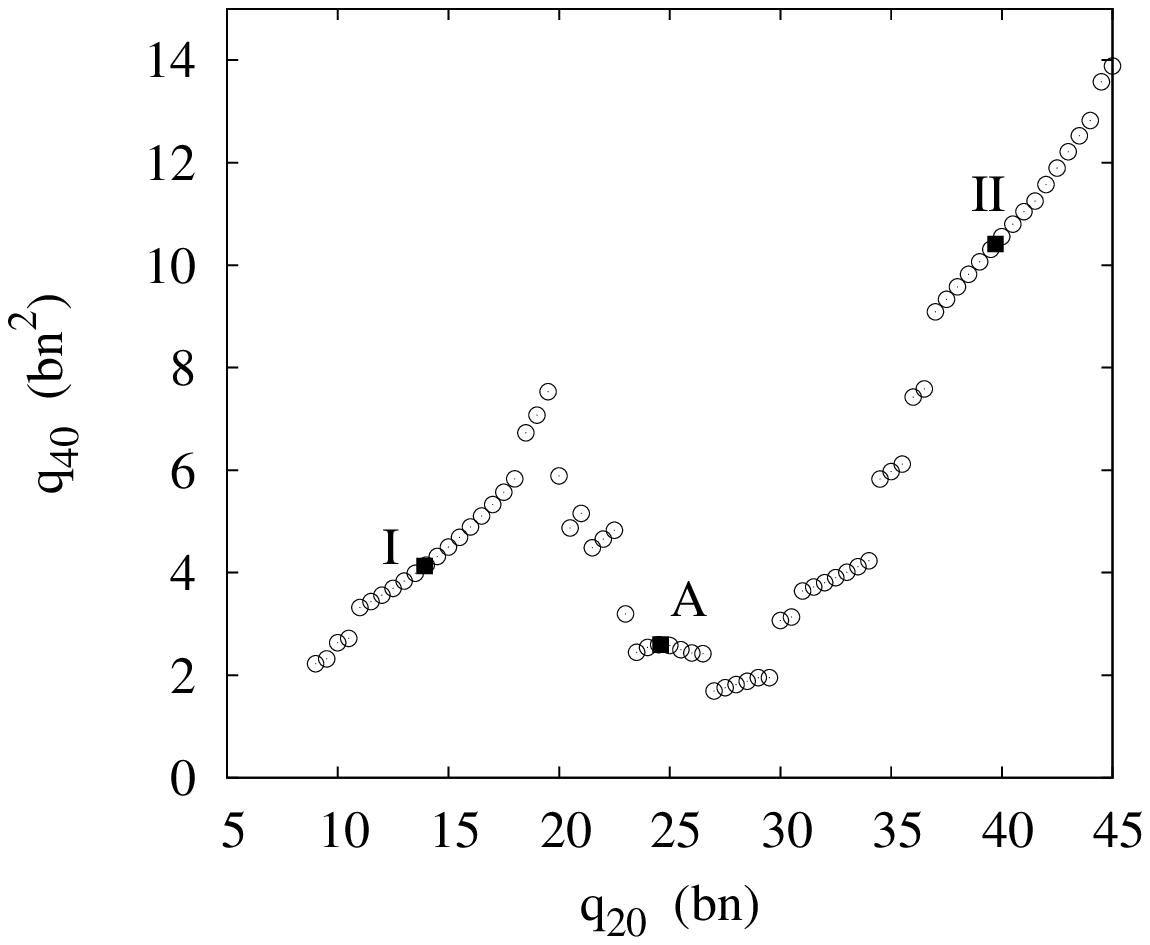}
}
\caption{Path from the first to second minimum in the $(q_{20},q_{40})$ plane.
The left-hand panel shows the HFB energies constrained by $q2$.  The
minima and intermediate maximum are marked by filled circles.
The right-hand panel shows the same path in the HF approximation.
The filled diamonds are the partition-constrained minima.  In principle,
either method should produce the same global minimum.
}
\label{u236-Evsq}
\end{figure} 
The lower curve was obtained by the GCM in the HFB
approximation, which includes pairing effects. 
The more jagged curve above
that is the GCM in the HF, ie. ignoring pairing
effects.   The HF configurations have
definite $K^\pi$ partitions that change abruptly along the curve.
The black dots are three 
configurations generated by the $K^\pi$-constrained  minimization,
with the partitions taken from the GCM results.
It is interesting to compare the $K^\pi$ partitions
found with the Gogny D1S energy function with other simpler ways for
building a many-particle wave function.  The simplest is the
Nilsson model mentioned earlier.  One finds that it gives exactly the 
same ground state $K^\pi$ partition as the DFT.  Another model simpler than the DFT is
the Finite-Range Liquid Drop Model \cite{mo09}.  Here
the orbital energies are computed in a deformed potential well, but the
shape of the well is determined by an energy functional based in part on the liquid drop model.
It also gives the same 
$K^\pi$ partition as the other models.
This suggests that the $K^\pi$ filling of the ground state is a quite
robust property of the mean-field wave function.  And it is certainly 
easier to specify, compared with the
many shape parameters needed to describe unambiguously configurations
obtained by the GCM.  However, at the A and the II points 
along the fission path the partition obtained with the Gogny functional
differs by one pair jump from the FRLD partition.  It is likely that
they are nearly degenerate, and would be mixed by the pairing interaction.

For dynamics, an important consideration is the number of
steps it takes to get from one configuration to another via the
two-particle interaction in the Hamiltonian.  This suggests a
distance measure between configurations as the number of pairing
jumps needed to get from one to the other.
Fig. \ref{u236-Evsq} shows a plot of the path
from configuration I to II via the barrier top at A, showing both the quadrupole
and hexadecapole coordinates of the intermediate configurations.  The
left-hand plot is the GCM in the HFB approximation, which is seen to
be smooth with very little structure.  The right-hand plot shows the 
same path in the HF approximation.  It is broken into a sequence of
segments, each segment having a specific $K^\pi$
partition.  Analyzing these segments, one can determine that it takes 
6 pair jumps to get
from I to A and 8 pair jumps to get from A to II.  But it would
wrong to concluded that it takes 14 pair jumps to get from  I and II.
In fact, examination of end-points reveals that the configuration 
II can be reached from I by changing the orbits of
only 6 pairs.  That of course requires surmounting a higher
potential barrier.  Which path is more important in the dynamics
is not obvious; one needs to know interaction matrix elements
between configurations as well as their energies to begin to
address this question.

\section{Conclusion}
The configuration-interaction method has
been very successful in light nuclei, and it should be 
possible, at least in a statistical way, to extend it to
heavy nuclei.  I have concentrated exclusively on the basis
of wave functions, advocating the use of DFT constrained by
$K\pi$ partitioning.  
This still leaves the interaction matrix elements to be 
determined.  The same problem exists in condensed-matter theory, and
there a hybrid approach 
has been quite successful \cite{sa16}.

It would take a large effort to carry out this program in nuclear
physics.  However, there are good reasons, rooted in experimental findings, 
to undertake that effort.  Empirically, there is strong evidence that the 
fissioning
nucleus is close to a statistical equilibrium near the scission point
\cite{ra11}.  The big open theoretical question is whether we
can explain that in terms of 
Hamiltonian dynamics with realistic interactions.

\begin{figure}[tb] 
\begin{center} 
\includegraphics[width=10cm]{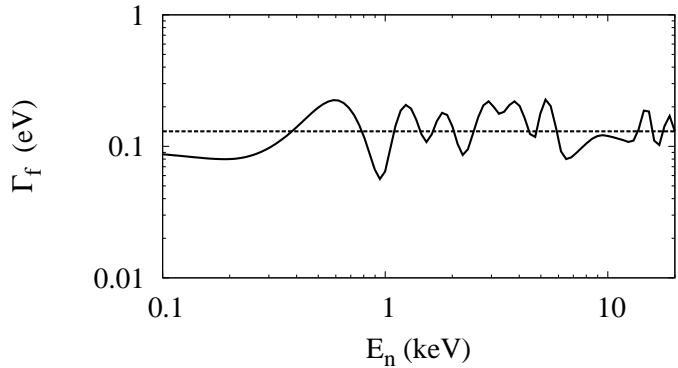}
\caption{Average fission width of the $J=4$ compound-nucleus states in $^{236}$U
as a function of excitation energy.  The dashed line is the predicted width
from the Bohr-Wheeler theory with one open channel. The solid curve is a 
fit to the experimental data in Ref. \cite{mo78}.
}
\label{moore} 
\end{center} 
\end{figure}

Fluctuation phenomena also remain unexplained.  Here are
two examples.  In 1978 Keyworth and collaborators
measured the fission cross sections for neutron-induced fission
of $^{235}$U separating out the individual angular momentum channels.
Besides the fluctuations associated with the compound nucleus states near
the waypoints I and II,  they saw fluctuations on 
much larger energy scale \cite{mo78}.  The data for one of the
angular channels is shown in Fig. \ref{moore}.  
It may be that the
discrete states near the barrier tops are responsible.  A discrete
basis would be very helpful here.

Another
old experiment exhibiting unexplained fluctuations is 
the measurement of angular distributions of fission fragments by
Huizenga, Loveland, and collaborators \cite{be69}.  The distributions 
very close
to threshold could be fitted with the usual theory, but on
examination of a more extended range they found:
``Further attempts to fit the data... were unsuccessful."

In summary, I think there is a good case 
for trying 
a discrete configuration approach to shape dynamics in heavy nuclei
as an alternative  to the GCM.  

%
\end{document}